\documentclass[aps,prc,twocolumn,nofootinbib,nopacs,nokeys,preprintnumbers,amsmath,amssy
mb,floats]{revtex4}
\usepackage[dvips]{graphicx}
\usepackage{subfigure}
\usepackage{amsmath}
\usepackage{amssymb}
\usepackage[rflt]{floatflt}

\begin{document}
\title{Gribov Poles in Diffractive Physics\footnote{Contribution to the proceedings of the \emph{Journ\'ee des Jeunes Chercheurs}, La Rochelles, France, December 2006.}}
\author{A. Dechambre}
\email{alice.dechambre@ulg.ac.be} \affiliation{Fundamental Interactions in Physics and Astrophysics, Universit\'e de Li\`ege, Belgium}

\date{\today}

\begin{abstract}
High-energy diffractive physics has several interests for theoreticians and experimentalists. We first remind the reader of the main characteristics of diffraction in particles physics and present the calculation of a two-gluon exchange in which non-perturbative effects are introduced via new singularities from confinement.
\end{abstract}

\maketitle

\section{Introduction}
Diffraction in high-energy Quantum Chromodynamics (QCD) is now actively studied from the theoretical and experimental points of view, largely because of the opening of the Large Hadron Collider~(LHC) in~2008. This new large accelerator will have an energy in the center-of-mass system of the order of 14~TeV and the detectors placed on the circumference will look for the Higgs boson, the last particle of the Standard Model still undiscovered. At the LHC, the protons move at the speed of light and scatter, producing from different processes a vast number of particles. The main practical interest of diffractive production is the rapidity gap between the produced hadrons. In diffractive collisions, a small number of particles are produced and these particles are well separated in the final phase space, without hadronic remnants so that the rare particles that can be produced decay without background.\\

\noindent In this paper, we first remind the reader of the main characteristics of diffraction in particles physics and of the general way to write a model of diffractive production of the Higgs boson. Next, we show why we need to worry about the infra-red (IR) region in diffractive models. The IR~region is affected by non-perturbative effects and one way to take some of them into account is to modify the gluon propagator of the theory giving it a finite value in the IR region. We will use this modified propagator in the last part of the paper to calculate two-gluon exchange.

\section{Diffractive Production}
\begin{figure}
\begin{center}\includegraphics[width=7cm,keepaspectratio]{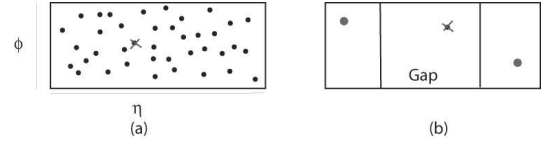}
\end{center}
\caption{\label{gap} Diagrammatic representation of the phase space after a proton/proton scattering in a collider. (a) Classical process, (b) Diffractive process.}
\end{figure}

Hadronic diffractive processes that take place at high energy~$\sqrt{s}$ are dominated by the exchange of a color singlet called pomeron with the quantum numbers of the vacuum. They are characterized by a large separation in rapidity of the final products and by few hadronic remnants. Practically, and contrarily to inelastic processes where each particle is hidden in numerous products of the reaction, the initial protons or their remnants are separated in the detectors and other particles centrally produced are isolated in the gap as we show graphically in Fig.~\ref{gap}.
From the experimental point of view, the interest is that, at LHC energy, about
50$\%$ of the total cross section will come from diffractive processes. The large rapidity gap allows an interesting method of measurement because it is indeed possible to measure precisely the momenta of the initial and final protons and to reconstruct the mass of the centrally produced resonance. From the theoretical point of view, it is possible to show that lowest-order QCD provides a good estimate of diffractive processes, i.e. the simplest models are in good agrement with data.

\subsection{Higgs diffractive production}
\begin{figure}
\begin{center}\includegraphics[width=5cm,keepaspectratio]{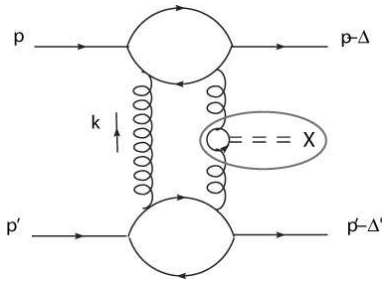}
\end{center}
\caption{\label{higgs}An example of diagram related to the diffractive cross section of the production of a Higgs boson in the high-energy limit.}
\end{figure}

Now, we can see why diffractive production of the Higgs boson
\begin{equation}
pp\to pHp
\end{equation}
 is so interesting for experimentalists and theoreticians. Furthermore, from the LEP results, the Higgs boson would have a mass close to 120 GeV and in this case, the largest branching ratio is to $b\bar{b}$. These quarks $b$ are practically impossible to tag in current detectors so that diffractive production gives us an alternative method of measurement from the missing mass as explained previously. Most of the mechanisms of diffractive Higgs production are similar to the one show in Fig.~\ref{higgs},
inspired by the original paper of Bialas and Landshoff~\cite{Bialas:1991wj}. Initial protons interact, exchanging two gluons and the final Higgs is then produced from a quarks loop.

\section{Introduction of non-perturbative effects}
Complex diffractive calculations can usually be separated in different pieces with the help of the $k_t$-factorization theorem that allows a separation between the kinematics of a process. This means that in the large-$s$ limit, we can treat separately the transverse kinematics typical of the exchange and the longitudinal kinematics of the impact factor. This is well-known in exchanges described by a BFKL ladder~\cite{BFKL} of which the simplest representation is two-gluon exchange. The problem is that the BFKL gluons can have a virtuality in all the energy domains, from ultra-violet to IR and that the processes that take place in the IR~region are characterized by a large coupling constant. These processes with large coupling can't be treated in perturbation theory~(pTheory) and they are modified by non-perturbative effects typical of the small virtuality or large distances. The question is then how to describe these effects, how to compute them, and see whether they modify the processes of production or if they respect the factorization theorem. The idea used in this contribution is to modify the gluon propagator in order to include non-perturbative effects directly in a perturbative calculation.\\

\begin{figure}
\begin{center}\hspace*{-2cm}\includegraphics[width=6.5cm,keepaspectratio]{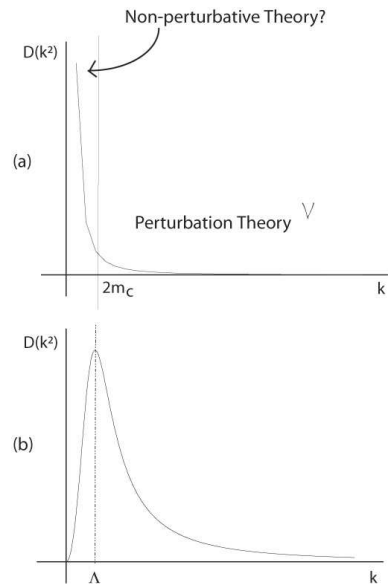}
\end{center}
\caption{\label{prop}(a) QCD gluon propagator, (b) Modified gluon propagator.}
\end{figure}

\noindent The QCD gluon propagator, in Feynman gauge, is usually written
\begin{equation}\label{propk}
D_{\mu\nu}(k^{2})=\frac{-ig_{\mu\nu}\delta^{ab}}{k^{2}+i\epsilon},
\end{equation}
where $a$ and $b$ are color indices and $k$ the momentum of the gluon. The shape of the propagator is drawn in Fig.~\ref{prop}.a,
it has the predicted form in the domain of large momentum~$k$ where pTheory can be used and diverges for small momentum. The divergence is due to a pole at~$k$=0 that can be interpreted as a particle propagating to large distances. But, as we know, gluons are confined so that we cannot trust the shape of the propagator~(\ref{propk}), something must happens in the IR~region in order to reproduce this confinement. According to Gribov~\cite{Gribov:1977wm} but also to several current works on Dyson-Schwinger equations~\cite{Bender:1994bv}, it is likely that the gluon propagator will be close to

\begin{equation}
\begin{split}
D_{\mu\nu}(k^{2})&=\frac{-ig_{\mu\nu}\delta^{ab}}{k^{2}+\frac{\Lambda^{4}}{k^{2}}}\\
&=-ig_{\mu\nu}\delta^{ab}\frac{k^{2}}{(k^{2}+i\Lambda^{2})(k^{2}+i\Lambda^{2})}.
\end{split}
\end{equation}
This new propagator, which we shall call the Gribov propagator, is not divergent but null at $k$=0 and the constant~$\Lambda$ is related to the energy scale where non-perturbative effects begin to appear. These effects do not modify the shape of the propagator when the momenta are large as shown in Fig.~\ref{prop}.b. Usually, one thinks that non-perturbative effects take place for virtuality below~2~or~3~GeV because beyond this scale, perturbation theory gives good results. Physically, the use of the Gribov propagator amounts to giving to the gluon an imaginary mass that includes non-perturbative effects via two new poles.

\subsection{Two-gluon exchange}
\begin{figure}
\begin{center}\includegraphics[width=6cm,keepaspectratio]{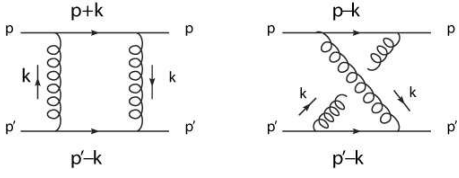}
\end{center}
\caption{\label{diaF}Lowest-order leading Feynman diagrams for the two-gluon exchange.}
\end{figure}

Now, we can study how the Gribov propagator influences the amplitude of a color-singlet exchange described by a two-gluon exchange. The lowest-order leading Feynman diagrams for this process are the direct and crossed diagrams at Fig.~\ref{diaF}.
The momenta~$k$ of the gluons are written on the light-cone and in term of the Sudakov variables,

\begin{equation}
k=\alpha p+\beta p'+ k_{t}.
\end{equation}
This representation decomposes the momenta of the gluons on the quadri-momenta of the incident particles and on the transverse direction. We can now understand $\alpha$ as the part of the momentum of the gluon in the direction of $p$, $\beta$ as the part in the direction of $p'$ and $k_t$ as the momentum in the direction transverse to the beam.\\

\begin{figure}
\begin{center}\includegraphics[width=8.6cm,keepaspectratio]{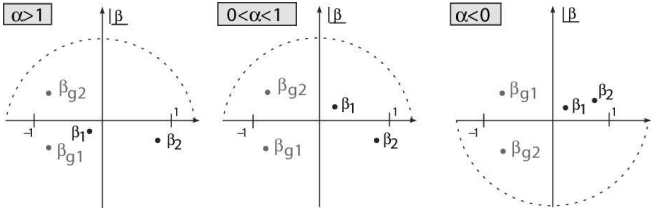}
\end{center}
\caption{\label{poleposition}Position of the poles in $\beta$ as a function of the value of~$\alpha$. $\beta_{i}$ ($i$=1,2) denoted the usual poles and $\beta_{gi}$ the new poles from the Gribov propagators.}
\end{figure}

\noindent We now consider the direct amplitude but it is clear that the method is identical for the crossed diagram. The amplitude for the direct process is given by,

\begin{equation}
\begin{split}
A_{d}&=(-ig_{s})^{4}C\int\frac{d^{2}\textbf{k}_{t}d\beta d\alpha}{(2\pi)^{4}} \quad \mathrm{U}(\textbf{k}_{t})\\
&\times \frac{(\alpha\beta s+\textbf{k}_{t}^{2})^{4}}{[(\alpha\beta s+\textbf{k}_{t}^{2}+i\Lambda^{2})(\alpha\beta
s+\textbf{k}_{t}^{2}-i\Lambda^{2})]^{2}},
\end{split}
\end{equation}
where $U(\textbf{k}_{t})$ is the contribution which includes only the quarks propagators. The new gluon propagator adds two double poles at

\begin{equation}
\begin{split}
\alpha\beta s+\textbf{k}_{t}^{2}+i\Lambda^{2}&=0 \quad \to \beta_{g1},\\
\alpha\beta s+\textbf{k}_{t}^{2}-i\Lambda^{2}&=0 \quad \to \beta_{g2}.
\end{split}
\end{equation}
The position of the poles in $\beta$ as a function of the value of $\alpha$ is shown in Fig.~\ref{poleposition}, where we have noted~$\beta_{i}$ ($i$=1,2) the usual poles and $\beta_{gi}$ the poles from Gribov propagators.
We only have to deal with the poles inside the integration contour that correspond to particles on their mass-shell, for example, we can see that the quark of momentum~$p+k$ is on its mass-shell when $\alpha$ is between~0~et~1. The $\beta_{g1}$-pole doesn't contribute but can be understood as a physical singularity corresponding to a decaying particle. On the other hand, the $\beta_{g2}$-pole brings a new contribution to the amplitude for all values of $\alpha$. This pole is a non-causal singularity and can be understood as a simulation of the confinement in the propagator. We now calculate the contribution of the poles to the amplitude by integrating along the contour according to the residue theorem, the sum of the direct and cross diagrams, in the high-energy limit is

\begin{equation}
\mathcal{M}^{s}_{g}=\frac{g_{s}^{4}}{4\pi}C\int\frac{d^{2}\textbf{k}_{t}}{(2\pi)^{2}}\frac{\mathrm{f}(\textbf{k}_{t}^{2},s/\Lambda^{2})}
{(i\Lambda^{2}-\textbf{k}_{t}^{2})^{3}\Lambda^{4}}.
\end{equation}
The amplitude is a function of $\textbf{k}_{t}$ and $s/\Lambda^{2}$ and we have once again to integrate it on~$\textbf{k}^{2}_{t}$. The results have unusual properties as a divergence when~$\textbf{k}_{t}$ tends to infinity. The divergence may get regulated by the introduction of an impact factor that would cut the integral. The main observation is that the answer is one order of $s$ lower than the usual contribution of the pole $\beta_{1}$,

\begin{equation}
\mathcal{M}^{s}_{u}=ig_{s}^{4}sC\int\frac{d^{2}\textbf{k}_{t}}{(2\pi)^{2}}\bigg{[}\frac{1}{\textbf{k}_{t}^{4}}\bigg{]}.
\end{equation}

\begin{figure}[b]
\begin{center}\includegraphics[width=8cm,keepaspectratio]{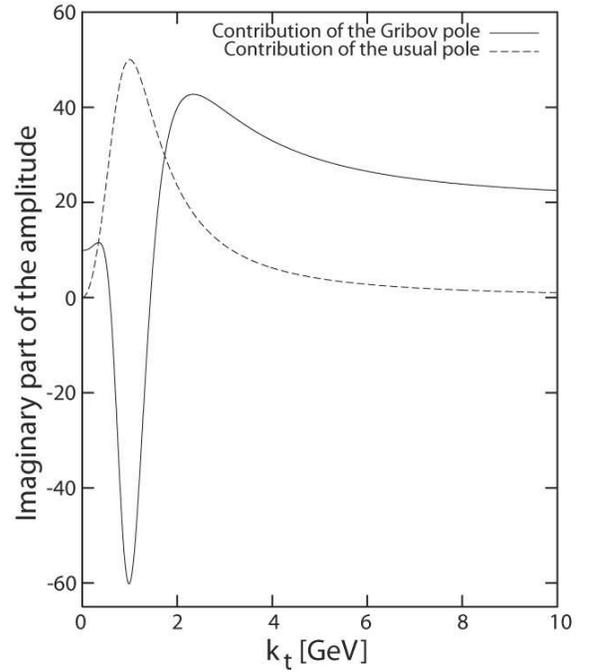}
\end{center}
\caption{\label{pimapimag}Imaginary part of the amplitude before the last integral as a function of $\textbf{k}^{2}_{t}$. Contribution of the Gribov pole $\beta_{g2}$ and of the usual pole $\beta_{1}$.}
\end{figure}
\subsection{First results}
In Fig.~\ref{pimapimag},
\begin{figure}
\begin{center}\includegraphics[width=8cm,keepaspectratio]{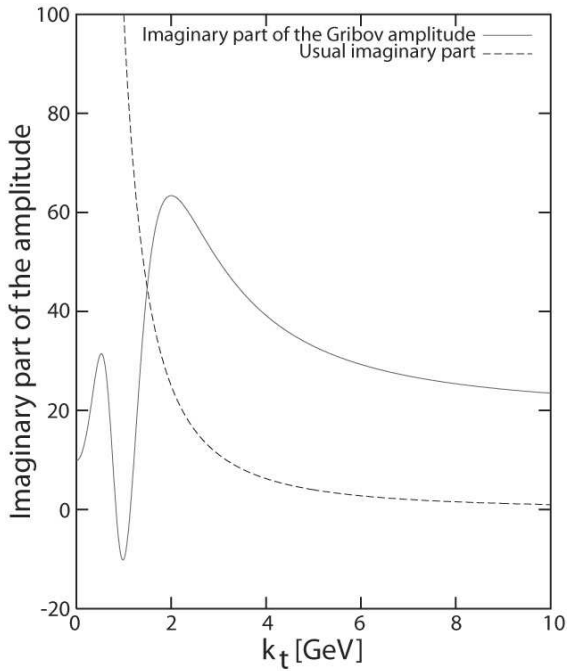}
\end{center}
\caption{\label{usuGpima}Imaginary part of the amplitude as a function of~$\textbf{k}^{2}_{t}$. With and without the modified gluon propagator.}
\end{figure}
we have drawn the contribution of the Gribov poles $\beta_{g2}$ as a function of~$\textbf{k}_{t}$, for a value of~$s=$100~GeV$^2$. We have fixed~$\Lambda=$1~GeV, i.e. an energy scale where we can be sure that non-perturbative effects are important. We first note that the Gribov contribution is smaller that the usual contribution and rapidly becomes a constant. If we now compare the imaginary part of the amplitude using the modified gluon propagator with the one using the QCD gluon propagator, we obtain Fig.~\ref{usuGpima}. The modified amplitude is negligible in the domain of interest.\\

The next step of the work is to introduce the impact factor, i.e. the coupling between the incident particles and the exchanged gluons.
\section{Conclusion}
In conclusion, we showed that the inclusion of non-perturbative singularities leads to effects which are probably small and don't influence the physics at the LHC. However, we can see that it would be interesting to study in details the influence of these effects on the amplitude and in particular if they always allow the use of the factorization theorem. We have to study a more complex model including the impact factor, firstly described by a quarks loop, and needed to avoid the divergences.

\end{document}